# Extreme temperatures characterization of $Al_{0.6}Ga_{0.4}N$ channel HEMT

Shuwen Xie[1,a)], Jacklyn Zhu[2], Swarnav Mukhopadhyay[1], Yi-Chen Liu[2], Shubhra Pasayat[1], Savannah Eisner[2], Chirag Gupta[1]

[1]*Department of Electrical and Computer Engineering, University of Wisconsin-Madison, Madison, WI, 53706 USA*

[2]*Department of Electrical Engineering, Columbia University, New York, NY, 10027 USA*

In this work, we fabricate and characterize ultra-wide bandgap $Al_{0.6}Ga_{0.4}N$ channel HEMTs with Pt/Au gate under a wide temperature range of 20 K to 650 K. The devices retained robust modulation capability throughout the measurement, with excellent on/off ratio of $9.7 \times 10^7$ and $6.4 \times 10^7$ under 20 K and 650 K. The DC output IV measurement was performed, with $I_{d,max}$ of a device with $L_{sd}$ = 9 µm measured to be 240 mA/mm and 120 mA/mm under 20 K and 650 K. In addition, low drain and leakage were observed in multiple devices, and the device robustness was evaluated under room temperature after the extreme temperature measurement with minimal degradation observed in $I_{d,max}$ and $R_{sheet}$. Moreover, transfer I-V characteristics of the $Al_{0.6}Ga_{0.4}N$ channel HEMTs have also been studied, where a relatively small $V_{th}$ shift was observed. TLM measurement and transconductance analysis both indicate the presence of weak temperature dependent scattering mechanisms in the device channel. The subthreshold swing (SS) was calculated and observed to increase parallelly to $SS_{ideal}$ under investigated temperature, indicates a robust gate structure. Finally, the on/off ratio and $I_g$ of the 9 µm $L_{sd}$ device were benchmarked and compared favorably against other reported results. While the results offer insight into the extreme temperature performance of AlGaN channel HEMT with Al composition higher than 50%, it also establishes AlGaN channel HEMT as a promising candidate for extreme temperature electronic applications.

Extreme temperature operation of electronic devices is critical for a wide range of applications. Deep drilling, aeronautical system, and gas turbines, for example, require robust high temperature capability, [1-2] while low temperature resilience is required for quantum computing, superconducting systems, as well as space exploration. [3-4] For high temperature applications, reliable operation of conventional Si based transistors are typically confined to temperatures below 300 °C. [5] Wide bandgap (WBG) and ultra-wide bandgap (UWBG) materials, with significantly larger bandgap energy, offer much lower intrinsic carriers under higher temperatures, thus enabling robust high temperature operation. [6-7] Among UWBG materials, AlGaN channel high-electron-mobility transistors (HEMTs) with high Al composition (Al > 50%) have emerged as one of the promising candidates for next generation high-frequency and high-power electronics. [8-13] With a bandgap larger than that of GaN, [14] high temperature resilient operation has also been demonstrated in AlGaN channel HEMTs in recent studies. [15-19] On the other extreme, power devices that rely on doped carrier drift layer, for example SiC MOSFET and Si IGBT, remain largely limited by carrier freeze-out and trapping dynamics for low temperature operation. [20-21] With insufficient thermal energy for dopant ionization under low temperature, the doped channel becomes more resistive and even semi-insulating, and large $V_{th}$ drift and severely degraded output

a) Authors to whom correspondence should be addressed: sxie75@wisc.edu

current have been observed in these devices [22-23]. In contrast, the sheet carrier concentration in 2-dimension electron gas (2DEG) channels of HEMT devices, including AlGaN channel HEMTs, is largely insensitive to low temperature. [24-25] The 2DEG channel is formed due to polarization charge and/or modulation doping, [26] making HEMTs largely immune to carrier freeze-out, thus inherently suitable for operation under cryogenic temperatures (< 120 K). [27] However, to date, the reports of AlGaN channel HEMTs (Al > 50%) performance under cryogenic temperatures remain scarce. [28-31] In this work, we fabricated $Al_{0.6}Ga_{0.4}N$ channel HEMTs with Pt/Au gate stack and characterized the devices from 20 K to 650 K under vacuum. Maximum current density was measured to be 240 mA/mm under 20 K ($V_{gs}$ = 0 V, $V_{ds}$ = 10 V) with source-to drain distance ($L_{sd}$) of 9 μm ($L_{gs}/L_g/L_{gd}$ = 2 μm/2 μm/5 μm). An excellent maximum on/off ratio was calculated to be > $10^7$ under temperature of 20 K and 650 K ($V_{gs}$ = 0 V, $V_{ds}$ = 8 V), respectively. Low drain and gate leakage were observed under the measured temperature range, indicating a robust gate structure. Transfer I-V of the device were also measured, and a relatively small $V_{th}$ shift was extracted. The temperature dependent sheet resistance and transconductance have been studied, suggesting enhanced phonon scattering under elevated temperature, and the presence of weak temperature dependent scattering mechanisms in the channel. The extracted subthreshold swing (SS) increased linearly with temperature, with a slope parallel to that of the thermal limit ($SS_{ideal}$). A minimal device degradation was observed after the extreme temperature operation.

The epitaxial structure of the $Al_{0.6}Ga_{0.4}N$ channel HEMT presented here was grown on a AlN/sapphire template with metal-organic chemical vapor deposition (MOCVD), shown in Fig. 1. From bottom to top, the structure comprises an $Al_{0.86}Ga_{0.14}N$ back barrier, a 200-nm $Al_{0.6}Ga_{0.4}N$ channel, a 1-nm AlN spacer, and a 24-nm Si:$Al_{0.81}Ga_{0.19}N$ barrier. On top of the barrier, a 50-nm $Al_xGa_{1-x}N$ (x: 0.82 -> 0.37) composition reverse graded (RG) layer was grown, followed by a 14-nm $n^{++}$ Si:$Al_{0.37}Ga_{0.63}N$ contact layer for contact resistance reduction. Details of the growth conditions were described elsewhere. [32] The device fabrication process began with Ti/Al/Ni/Au (20/120/30/50 nm) ohmic metal stack deposition using e-beam evaporation, followed by rapid thermal annealing (RTA) under 1000°C in a $N_2$ ambient. Next, a time-dependent $BCl_3/Cl_2$ based low power reactive ion etching (RIE) was performed to remove the RG layer in the access region. The device was then isolated using a $BCl_3/Cl_2$ based RIE with RF power of 15 W. Finally, gate metal was formed using lift-off process following Pt/Au (50/100 nm) deposition using e-beam evaporation.

The AlGaN channel HEMTs were characterized using a Lake Shore Cryotronics CRX-6.5K cryogenic probe station. The DC output and transfer I-V characteristics of the devices were measured from 20 K to 650 K in a vacuum chamber. The chamber temperature was first lowered from room temperature to 20 K, then raised to 50 K, and subsequently increased in steps of 100 K up to 650 K. The temperature was stabilized at the set points and held for 1 h to impose thermal stress on the devices while characterizations were performed.

Transfer length method (TLM) measurement was performed with contact spacings varying from 2.5 um to 40 um under 20 K to 650 K. Fig.2 a)-b) shows the measured resistance-spacing relationship and linear extrapolated sheet resistance ($R_{sheet}$) under different temperatures. The extracted $R_{sheet}$ increased by 81% from 5510 Ω/□ at 20 K to 9974 Ω/□ at 650 K. This temperature dependent increase of $R_{sheet}$ could be attributed to a degraded electron mobility due to increased optical phonon scattering under higher temperature [6, 28]. On the other hand, relatively stable $R_{sheet}$ values were observed under low temperature, which could result from the domination of temperature-independent interface roughness scattering and alloy scattering as electron-phonon interaction reduced under lower temperature. [28] To verify the robustness of the ohmic metal contact, TLM measurement was performed again under room temperature before and after the 9-hour duration temperature dependent measurement. The comparisons were shown in Fig 2. c)-d). The extracted contact resistance ($R_c$) increased from 3.3 Ω·mm to 3.6 Ω·mm, respectively. After the extreme temperature stress, the extracted $R_{sheet}$ showed a minimal increase from 5513 Ω/□ to 5619 Ω/□, correspondingly a degradation of 1.9%. While identical $R_{sheet}$ were expected to be extracted before and after testing, the slight variation might arise from data fitting as well as slight degradation of ohmic contact.

The temperature dependent DC output I-V characterization was performed on a $Al_{0.6}Ga_{0.4}N$ channel HEMT with $L_{gs}/L_g/L_{gd}$ of 2 µm/2 µm/5 µm. Fig. 3 a)-b) shows the output current density at $V_{gs}$ = 0 V under various temperatures. With elevated temperature, the output characteristic exhibited an overall decline under the $V_{ds}$ measurement range. Quantitatively, the maximum drain current density ($I_{d,max}$), evaluated at $V_{gs}$ = 0 V and $V_{ds}$ = 10 V, drops by 50% from 240 mA/mm at 20 K to 120 mA/mm at 650 K. The evolution of current density corroborated the temperature-dependent behavior of the $R_{sheet}$ as described above, indicating intensified optical phonon scattering as temperature increased. [6, 28] The complete DC output I-V characteristics under 20 K, 250 K and 650 K with $V_{gs}$ from 0 V to -10 V were shown in Fig. 3 a), with the device retaining robust channel modulation capability throughout the measurement. Meanwhile, the drain leakage current ($I_{d,off}$) measured under $V_{gs}$ = -10 V and $V_{ds}$ = 10 V remained below $5 \times 10^{-8}$ A/mm throughout the measured temperature range, which was approximating the lower current measurement limit of the tool, indicating a robust gate structure under these extreme temperatures. The DC output I-V of the device was characterized again under room temperature after the temperature dependent measurements, with comparison shown in Fig. 3 c). A minimal degradation of 4% was observed in $I_{d,max}$, while $I_{d,off}$ remained below $5 \times 10^{-8}$ A/mm. To verify the consistency of the output I-V measurement, five devices with various $L_{gs}/L_g/L_{gd}$ were subsequently characterized, summarized in Table 1. The temperature-dependent $I_{d,max}$ ($V_{gs}$ = 0 V, $V_{ds}$ = 10 V), and on-resistance ($R_{on}$) ($V_{gs}$ = 0 V, $V_{ds}$ = 1 V) of the were shown in Fig.4 a)-b). Low $I_{d,off}$ were measured for the five devices during the temperature dependent measurement, shown in Fig. 4 c). Regardless of the device dimensions, an $I_{d,max}$ degradation was observed as temperature increased, in agreement with the previously described temperature dependent evolution of $I_{d,max}$. On the other hand, the temperature dependence of the extracted $R_{on}$ was consistent with that of $R_{sheet}$ described earlier. These results suggest that the temperature dependent $R_{on}$ variations in the measured devices were primarily governed by $R_{sheet}$ variations.

The transfer I-V characteristics were subsequently measured for the device with $L_{gs}/L_g/L_{gd}$ of 2 µm/2 µm/5 µm. The measurement was performed under $V_{ds}$ of 0.5 V and 8 V from 20 K to 650 K, shown in Fig. 5 a)-b). The inset of Fig. 5 a) compares the room temperature transfer I–V characteristics of the same device before and after measurement, with negligible degradation observed. The temperature dependent gate leakage ($I_{g,off}$) measured under $V_{gs}$ = -10V and $V_{ds}$ = 8 V, shown in Fig. 5 c). The $I_{g,off}$ remained below $5 \times 10^{-8}$ A/mm throughout the temperature range, corroborating a robust gate stack under measured extreme temperatures. The on/off ratio of the device was measured to be $9.7 \times 10^7$ under 20 K, and $6.4 \times 10^7$ under 650 K, shown in Fig. 5 d). The on/off ratio of all the five devices (as described in Table I) was shown in the inset of Fig. 5 d). The measured on/off ratio remained relatively stable from 20 K to 650 K, with fluctuations dominated by off-state noise current originated from the measurement setup. It should be noted that the measured off-state current level is approximating the noise level of the measurement unit. With lower current measurement limit, the extract on/off ratio was expected to further increase.

Temperature dependent transconductance ($g_m$) and threshold voltage ($V_{th}$) of the device were extracted at $V_{ds}$ of 0.5 V, as shown in Fig. 6 a) and Fig. 6 b). The threshold voltage ($V_{th}$) of the devices was calculated using peak transconductance extrapolation method, [15, 33] where $V_{th} = V_{gs,peak} - I_{d,peak}/g_{m,max}$. Here, $V_{gs,peak}$ and $I_{d,peak}$ denote $V_{gs}$ and $I_d$ value at $g_{m,max}$. $V_{gs,peak}$ and $V_{th}$ decreased as temperature increased from 20 K to 350 K. It is hypothesized that higher $V_{th}$ under low temperature could be a result of trap freeze-out at the Schottky gate/AlGaN barrier interface. [20] While the presence of potential interface traps may lead to lowered Schottky barrier height through trap-assisted tunneling, trap freeze-out mitigates trap-assisted tunneling. Consequently, the Schottky barrier height was restored under low temperature with higher $V_{th}$ expected. On the other hand, a relatively stable $V_{th}$ was observed beyond 290 K, with variation within 0.2 V. Further investigation will be needed to further analyze the mechanisms affecting the $V_{th}$ drift. It should be noted that, compared with devices implemented with impurity doped channel, for example Si and SiC, the AlGaN channel HEMT showed lower $V_{th}$ drift as temperature decreased, [22-23] highlighting the inherent low temperature advantage of the 2DEG channel.

Temperature dependency was also observed in the value of maximum $g_m$ ($g_{m,max}$), shown in Fig. 6 c). As a low $V_{ds} << V_{gs} - V_{th}$ was applied, the AlGaN HEMT was operating in linear regime. Considering $I_d \approx W/L \cdot C_{ox}\, \mu_{eff}\, V_{ds} \cdot (V_{gs} - V_{th})$, where $W$ is gate width, $L$ is gate length, $C_{ox}$ is gate capacitance, $V_{th}$ is threshold voltage. Using $\partial I_d / \partial V_{gs}$, the relationship between $g_m$ and $\mu_{eff}$ can be deduced, where $g_m \propto \mu_{eff}$. To investigate the relationship between $\mu_{eff}$ and temperature, the $g_{m,max}$ data was fitted using $g_{m,max} = A \cdot T^{-\alpha}$ from 250 K to 650 K where optical phonons strongly interact with 2DEG electrons, shown in Fig. 6 d), where $A$ is constant, $\alpha$ is temperature coefficient. [34] The average value of $\alpha$ was extracted to be 0.9, yield $g_{m,max} \propto T^{-0.9}$, thus $\mu_{eff} \propto T^{-0.9}$. The temperature dependence of $W$, $L$, and $C_{ox}$ was neglected due to their relatively small variations over the investigated temperature range. It should be noted that, while a typical $\mu_{eff} \propto T^{-1.8}$–$T^{-2.6}$ [6] was observed in GaN HEMT, the $\mu_{eff} \propto T^{-0.9}$ in the AlGaN channel HEMT showed weaker temperature dependency. While optical phonon scattering still plays a dominating role under these temperatures, this indicates the presence of scattering mechanisms that are insensitive to temperature variations in the 2DEG channel, for example, alloy scattering and interfacial roughness scattering. The relatively temperature insensitive nature of AlGaN channel HEMT could be beneficial for the stable operation of analog circuits under extreme temperatures.

The temperature dependent subthreshold swing (SS) of the device was shown in Fig. 6 e). Here, SS was extracted as $SS\ (mV/dec) = d\,V_{gs} / d\,(\log_{10} I_d) \times 10^3$. The ideal subthreshold slope thermal limit, $SS_{ideal} = \ln(10)*kT/q$, was also plotted in Fig. 6 e) for reference, where $k$ is Boltzmann constant, $q$ is elementary charge. As temperature increases, SS was observed to increase linearly, parallel to the slope of $SS_{ideal}$. [35] This non-diverging SS increase compared with $SS_{ideal}$ indicates that the device was not suffering from thermal failure within the investigated temperature range.

The temperature vs. on/off ratio result reported in this work was benchmarked against other reports of WBG/UWBG transistors in Fig 7. a). The temperature dependent $I_{g,off}$ vs. on/off ratio result was benchmarked against other reported AlGaN channel HEMT reports, shown in Fig 7. b). While our AlGaN channel HEMT showed on/off ratio higher than SiC, [36] AlN, [37-38] $Ga_2O_3$ [39-41] and diamond [42-44] based devices, it also compares favorably with state-of-the-art GaN [20, 25, 45-54] and AlGaN [15, 16, 18, 29, 55] channel HEMTs with low leakage under the measured temperature range.

In conclusion, we demonstrated the operation of $Al_{0.6}Ga_{0.4}N$ channel HEMT with Pt/Au gate from 20 K to 650 K. $I_{d,max}$ of the device was measured to be 240 mA/mm and 120 mA/mm under 20 K and 650 K. The device retained modulation capability throughout the measurement, with on/off ratio measured to be $9.7 \times 10^7$ under 20 K, and $6.4 \times 10^7$ under 650 K. A low $I_{d,off}$ and $I_{g,off}$ below $5 \times 10^{-8}$ A/mm under the measured temperature range, and the robustness of the device was verified after the extreme temperature measurement with minimal degradation observed in $I_{d,max}$ and $R_{sheet}$. Transfer I-V has been studied, and a relatively small $V_{th}$ shift was observed. TLM and transconductance analysis indicated the presence of weak temperature dependent scattering mechanisms in the device channel. Linear increase of SS parallel to $SS_{ideal}$ indicated a robust gate structure, and the benchmark of the extracted on/off ratio and $I_g$ demonstrated favorable comparison against other reported results. While the $Al_{0.6}Ga_{0.4}N$ channel HEMT demonstrated promising performance under extreme temperature, the operation limit of the gate structure as well as ohmic contact remains to be explored. Further investigation into the influence of epitaxial growth, device passivation, choice gate metal stack will be needed to further improve the performance as well as stability under extreme temperatures.

ACKNOWLEDGEMENT

We acknowledge the University of Wisconsin-Madison Nanoscale fabrication center for providing research environment. This material is based upon work supported by the Air Force Office of Scientific Research under award number FA9550-25-C-B010 in the amount of $144,500.

DATA AVAILABILITY

The data that support the findings of this study are available from the corresponding author upon reasonable request.

# Figures

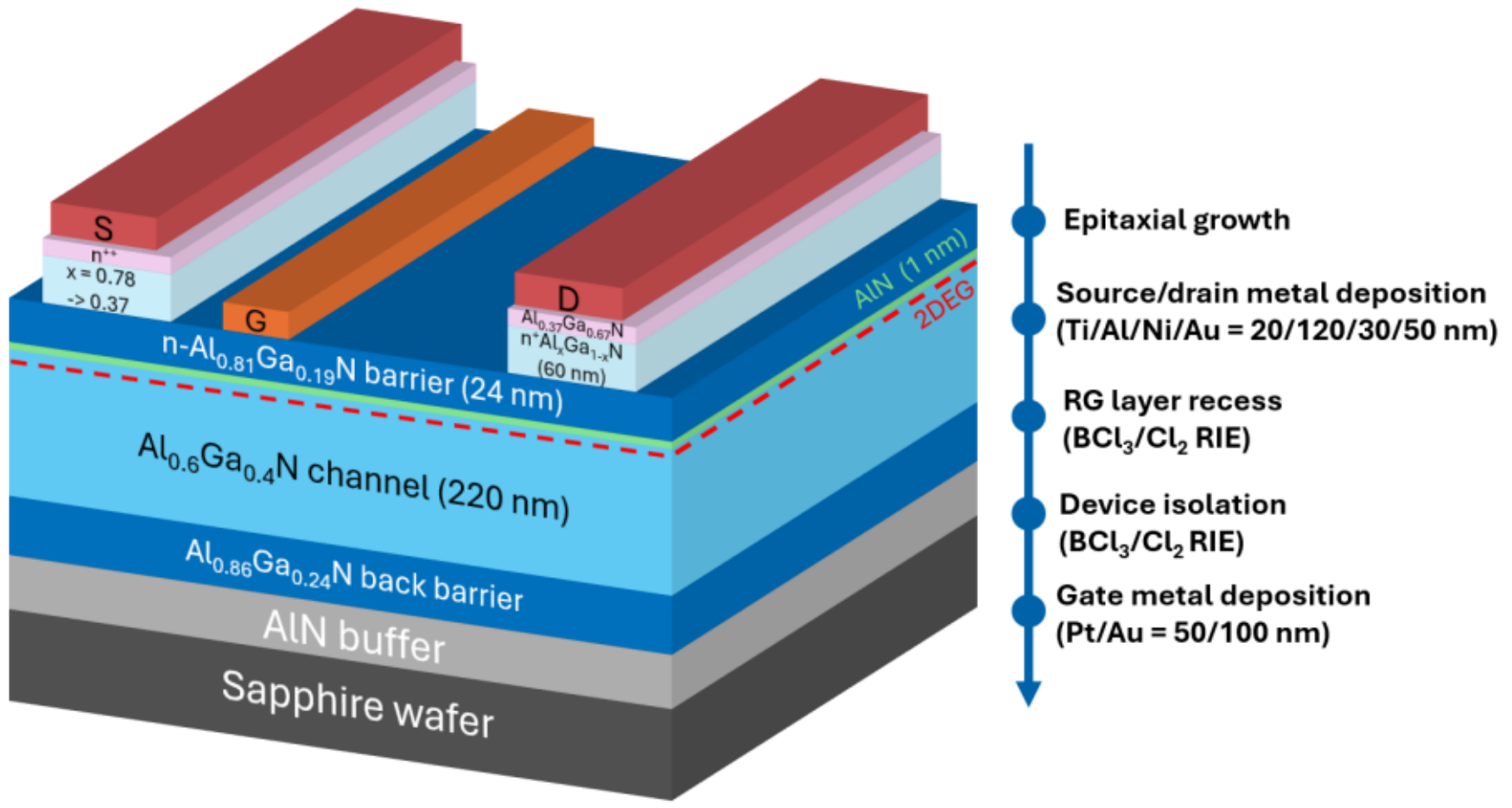

FIG. 1. Device structure and fabrication process flow of the $Al_{0.81}Ga_{0.19}N/Al_{0.6}Ga_{0.4}N$ HEMT.

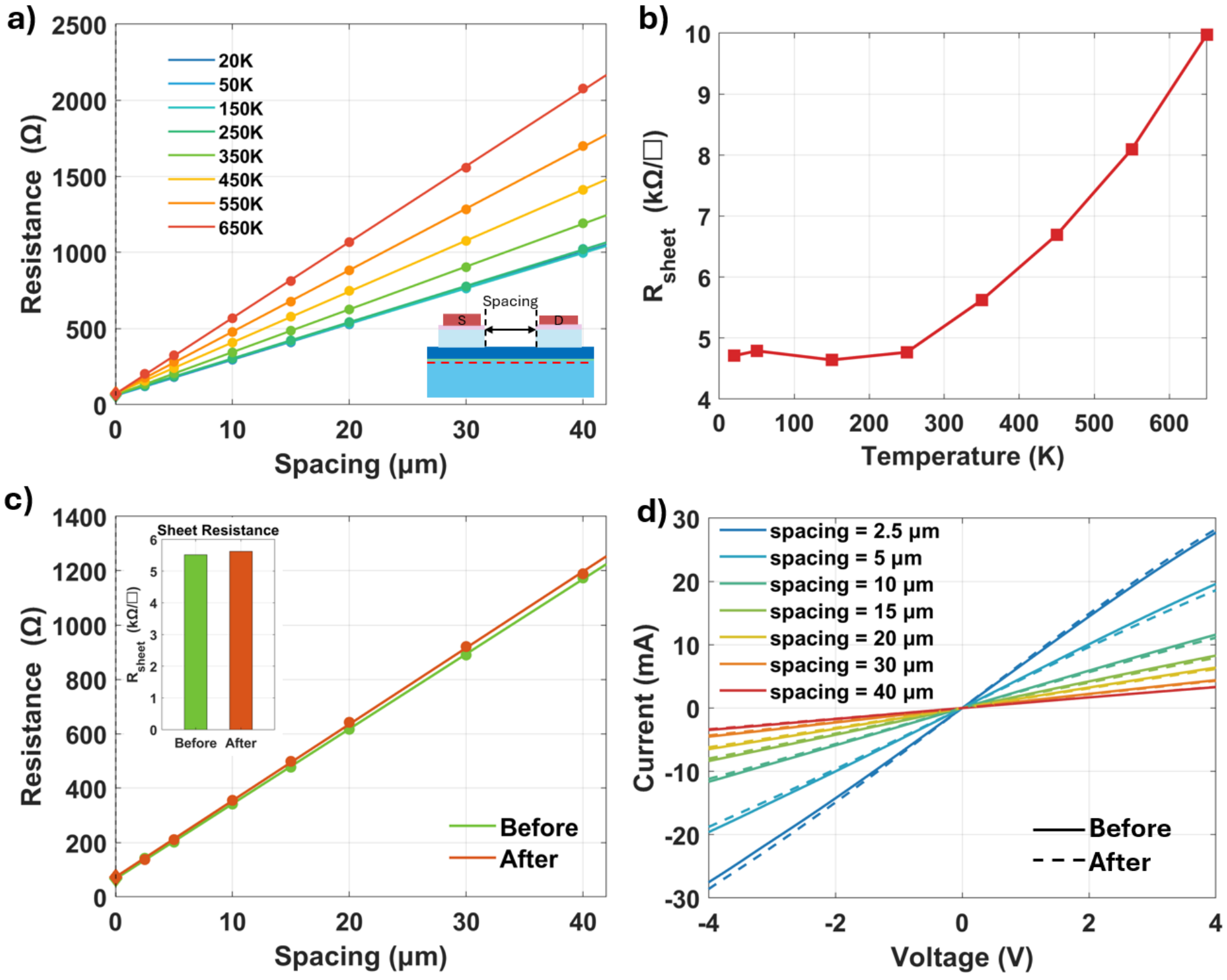

FIG. 2. a) Linear extrapolated TLM measurement results from 20 K to 650 K, and b) extracted temperature dependent $R_{sheet}$. c) Linear extrapolated and d) TLM measurement results before and after extreme temperature stress.

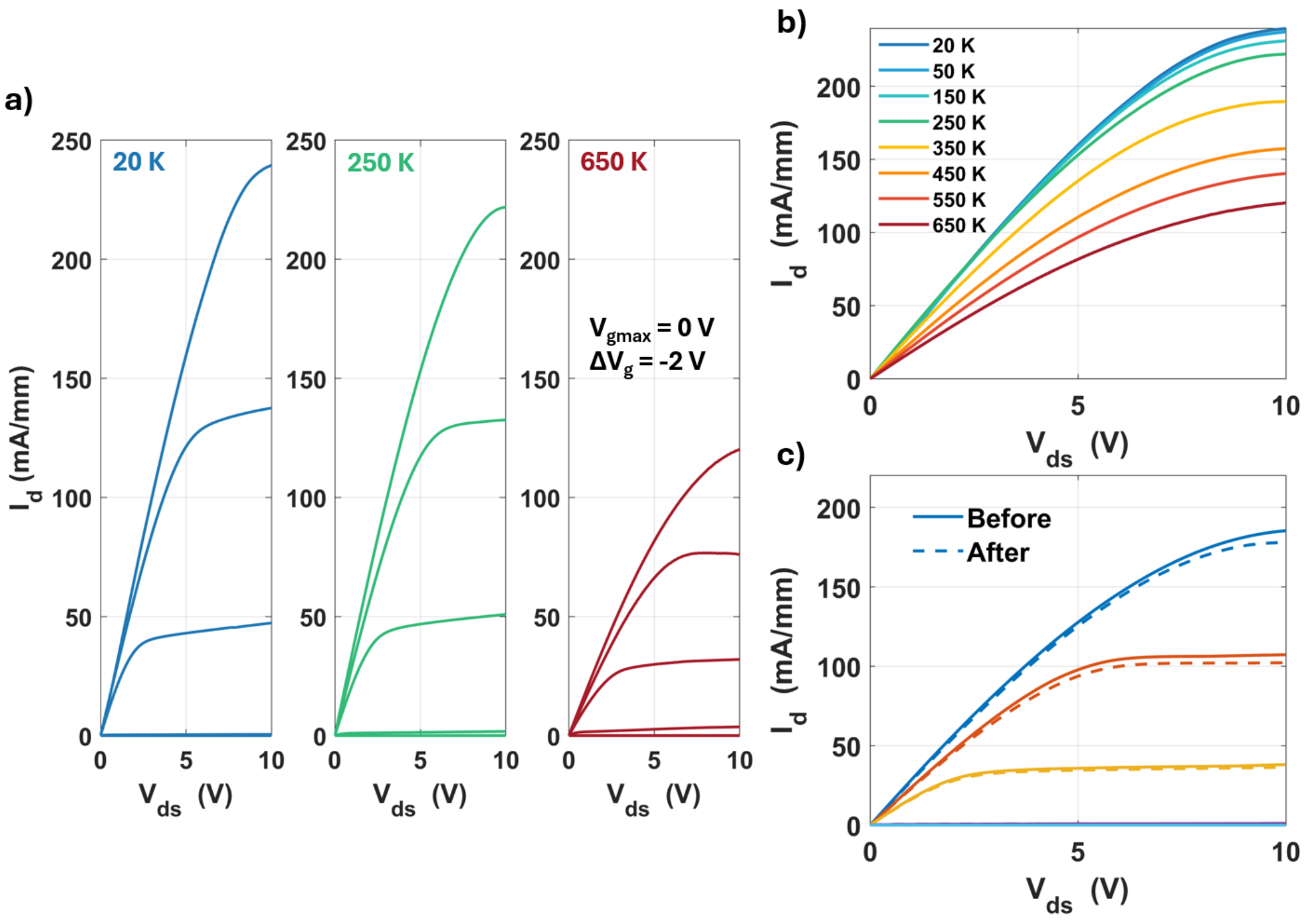

FIG. 3. a) Complete DC output I-V characteristics of a device with $L_{sd}$ of 9 μm under 20 K, 250 K and 650 K. b) DC output I-V characteristics of the 9 μm $L_{sd}$ device from 20 K to 650 K under $V_g$ = 0 V. c) DC output comparison before and after extreme temperature stress.

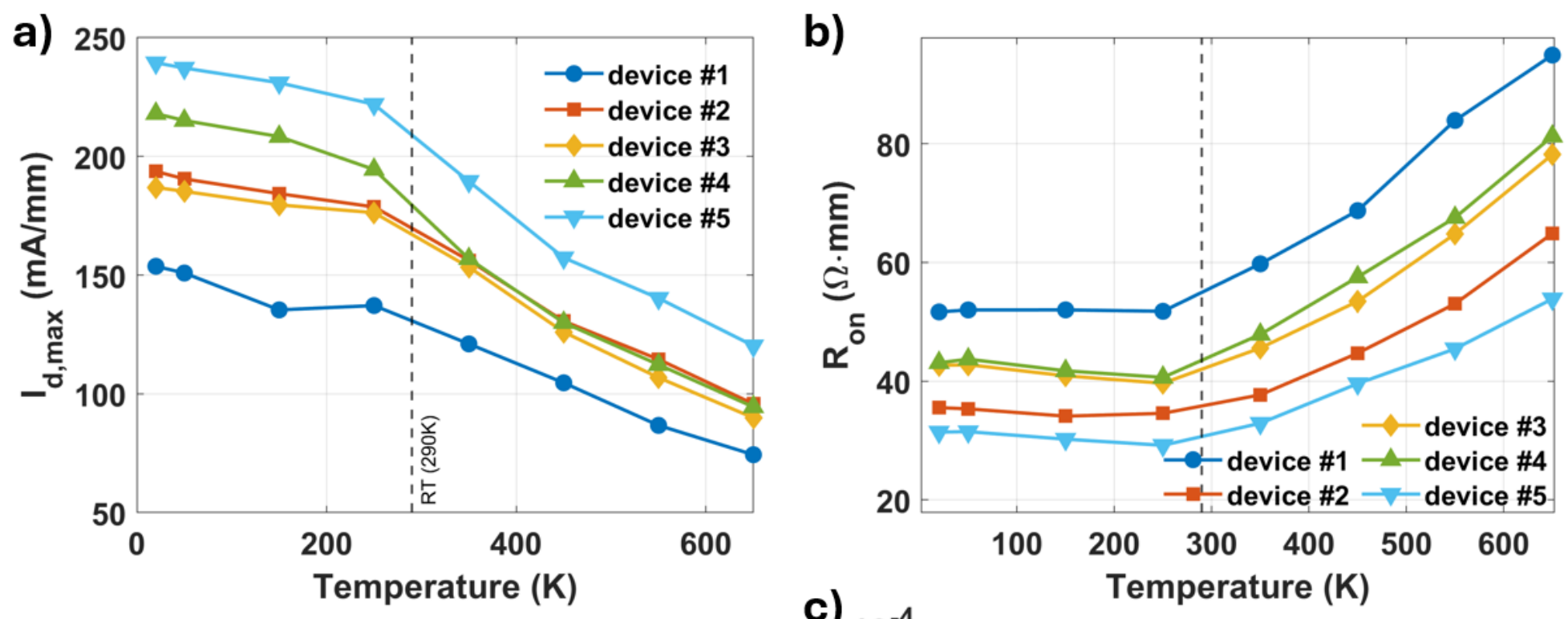

| Device # | $L_g$ (μm) | $L_{gs}$ (μm) | $L_{gd}$ (μm) |
|---|---|---|---|
| 1 | 2.5 | 2.5 | 10 |
| 2 | 2 | 2.5 | 5 |
| 3 | 2.5 | 2 | 7.5 |
| 4 | 2 | 2 | 10 |
| 5 | 2 | 2 | 5 |

**Tabel I**

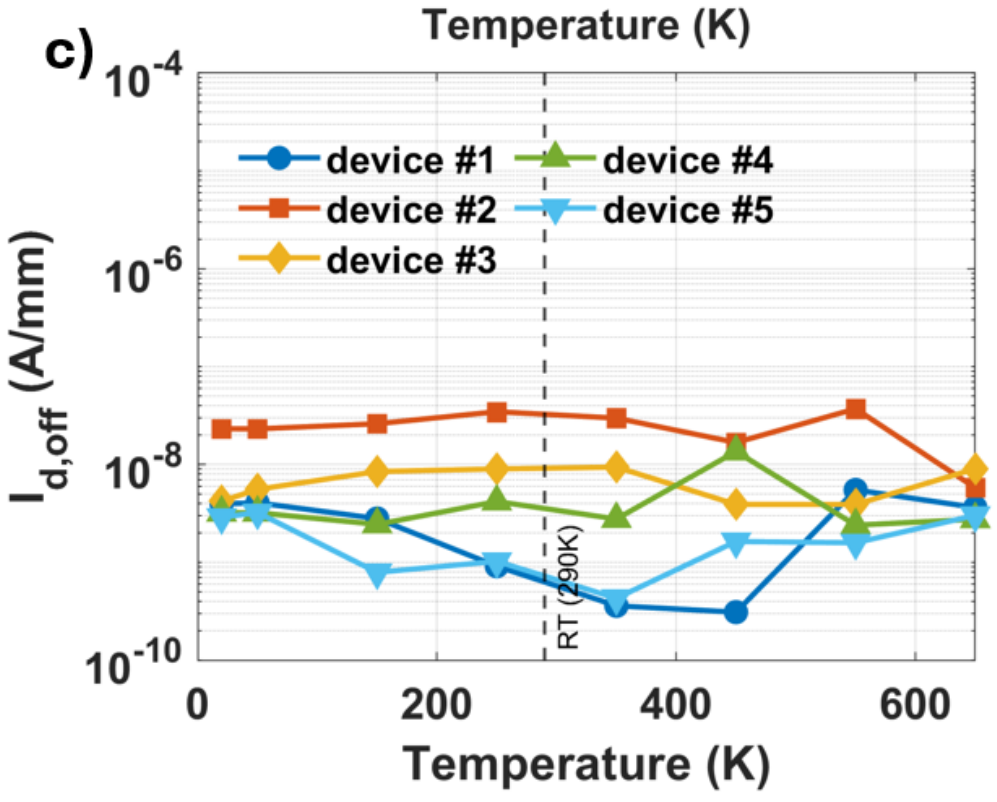

FIG. 4. Temperature dependent a) $I_{d,max}$, b) $R_{on}$, and c) $I_{d,off}$ of five devices with different $L_{gs}/L_g/L_{gd}$, summarized in Table I.

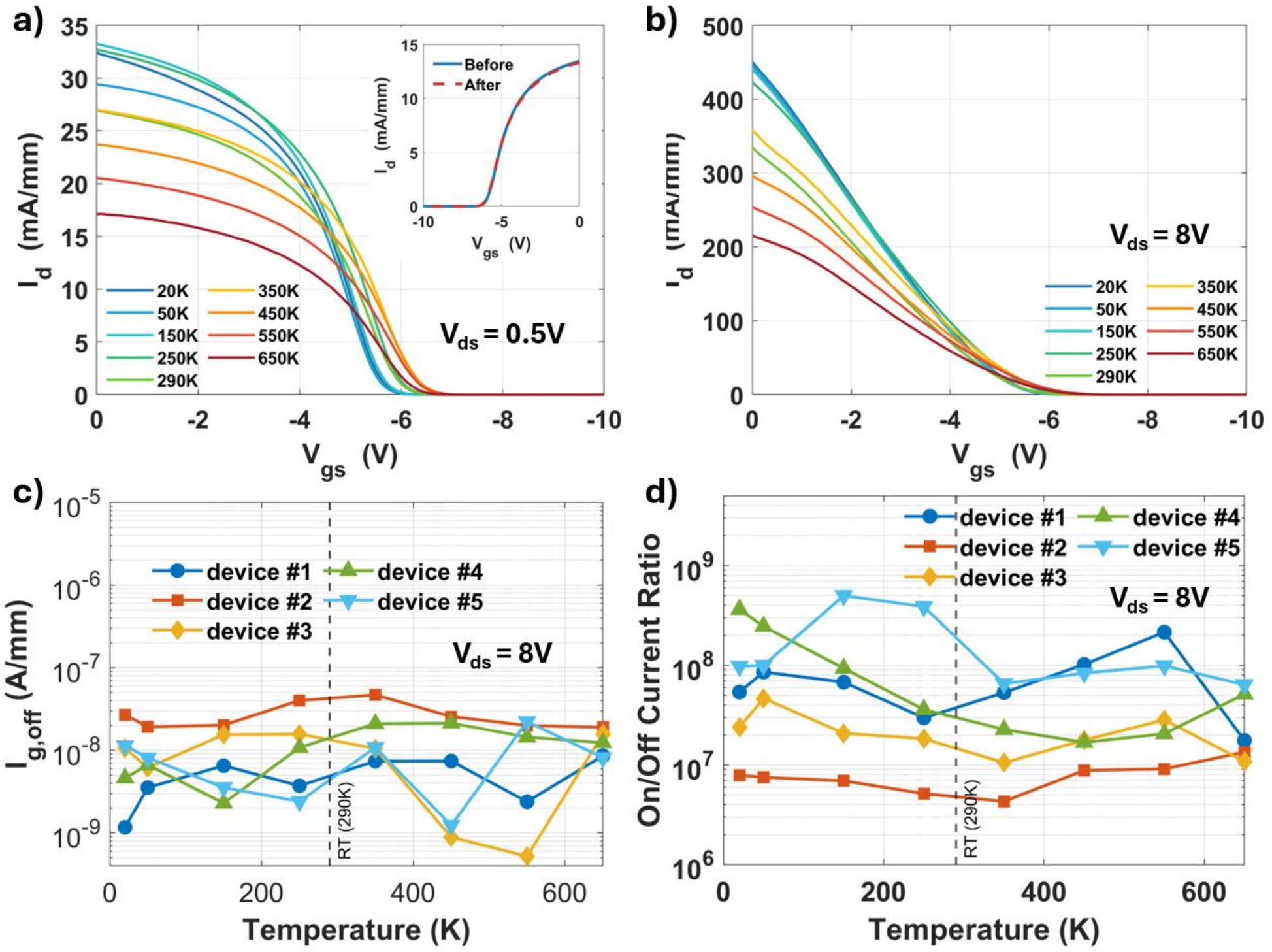

FIG. 5. Temperature dependent transfer I-V characteristics of the 9 μm $L_{sd}$ device under a) $V_{ds}$ = 0.5 V, and b) $V_{ds}$ = 8 V. Inset of a) compares transfer I-V before and after extreme temperature stress. c) $I_{g,off}$ and d) on/off ratio of five devices with different $L_{gs}/L_g/L_{gd}$ under $V_{ds}$ = 8 V.

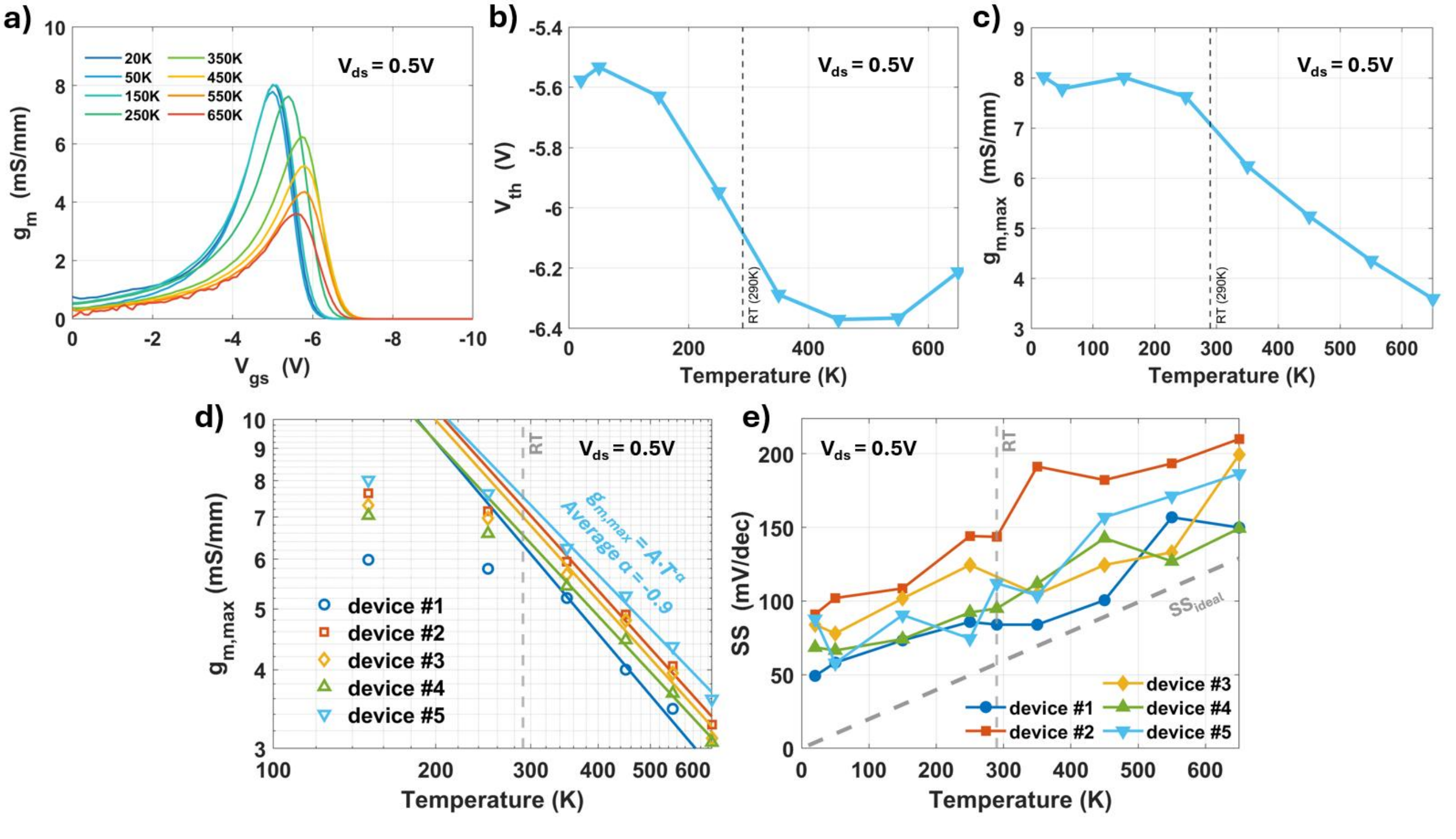

FIG. 6. Temperature dependent a) $g_m$, b) $V_{th}$, and c) $g_{m,max}$ of the 9 μm $L_{sd}$ device. d) $g_{m,max}$ data from 350 K to 650 K was fitted with $g_{m,max} = A \cdot T^{-\alpha}$, with average $\alpha$ extracted to be -0.9. c) Temperature dependent SS of five devices with different $L_{gs}/L_g/L_{gd}$, compared with $SS_{ideal}$.

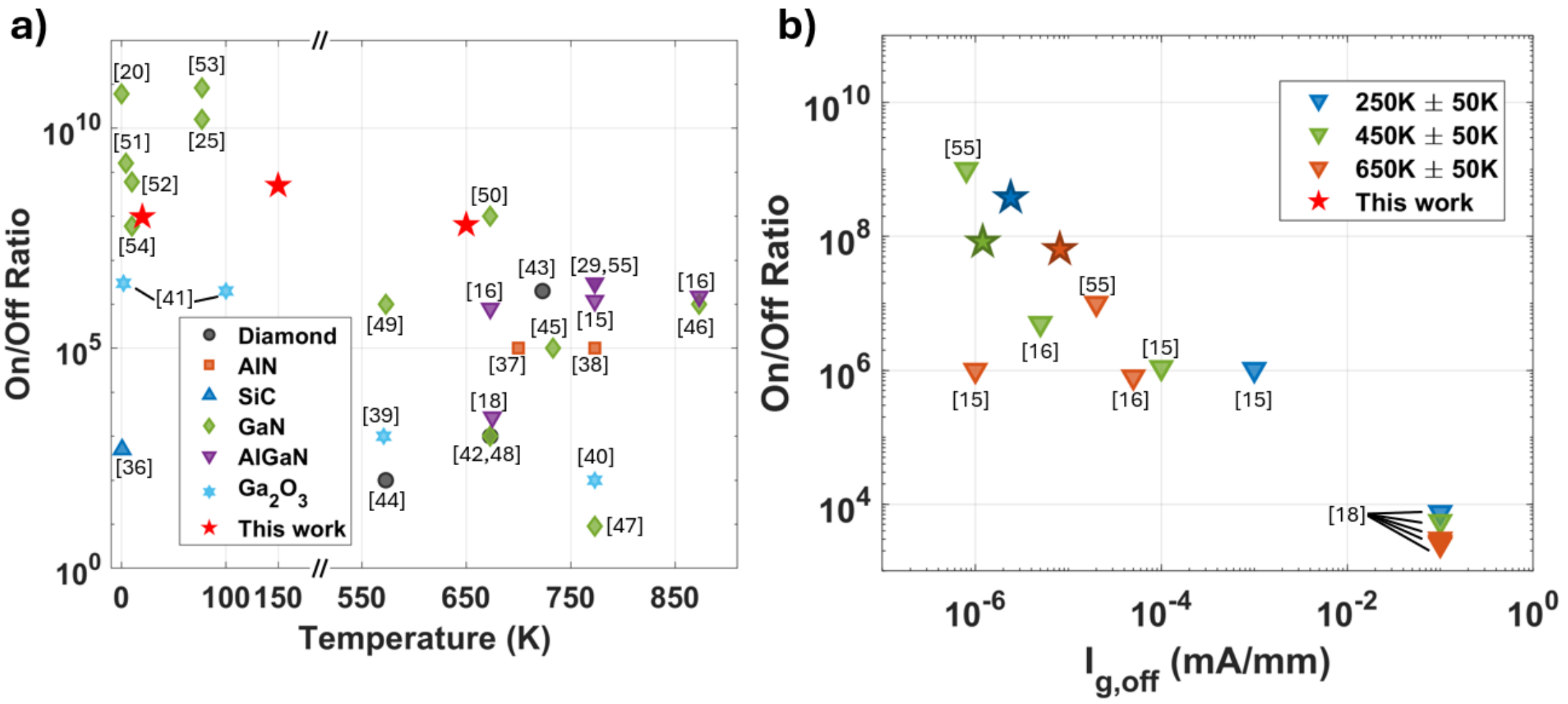

FIG. 7. a) Benchmark of temperature vs. on/off ratio results of WBG/UWBG transistors, and b) $I_{g,off}$ vs. on/off ratio results of AlGaN channel HEMTs.